\documentclass{ws-procs9x6}

\begin{document}

\title{Heavy quark production at HERA\\
 with BFKL and CCFM dynamics\footnote{\uppercase{T}his work is supported by the
\uppercase{R}oyal \uppercase{S}wedish \uppercase{A}cademy of 
\uppercase{S}cience. }}

\author{S.~P. BARANOV}

\address{Lebedev Institute of Physics\\
Leninsky prosp. 53, 
Moscow 117924, Russia\\ 
E-mail: baranov@sci.lebedev.ru}

\author{H. JUNG, L. J\"ONSSON}

\address{Department of Elementary Particle Physics, \\ 
Lund University, 
Box 118 SE 22100 Lund,Sweden\\
E-mail: jung@mail.desy.de \\
E-mail: leif.jonsson@quark.lu.se}

\author{S. PADHI
\footnote{\uppercase{W}ork partially
supported  by the \uppercase{N}atural
\uppercase{S}cience and  \uppercase{E}ngineering \uppercase{R}esearch
\uppercase{C}ouncil of \uppercase{C}anada.}}

\address{Department Physics, McGill University, \\
Montreal, Quebec, Canada, H3A 2T8\\
E-mail: Sanjay.Padhi@desy.de} 

\author{N.~P. ZOTOV
\footnote{\uppercase{W}ork partially
supported  by the \uppercase{R}ussian
\uppercase{F}ound  \uppercase{B}asic \uppercase{R}esearch
\uppercase{G}rant 02-02-17513.}}

\address{SINP, Moscow State University, \\
Moscow 119992, Russia\\
E-mail: zotov@theory.sinp.msu.ru} 
               
\maketitle

\abstracts{
In the framework of the semi-hard ($k_t-$factorization) approach, we
analyze the various charm production processes in the kinematic region
covered by the HERA experiments.}

\section{Introduction}
The present note is short version of our paper \cite{bjjpz} where we have
attempted a systematic comparison
of the theoretical predictions of the $k_t$-factorization approach
\cite{Gribov,Collins,CCH}
 with experimental data
regarding the charm production processes
at HERA.

The production of
 open-flavored $c \bar c -$ pairs in $ep$-collisions
is described in terms of the photon-gluon fusion mechanism.
A generalization of the usual parton model to the $k_t$-factorization
approach implies two essential steps. These are the
introduction of
unintegrated gluon distributions and the modification of the gluon spin
density matrix in the parton-level matrix elements.
The hard scattering cross section for a boson gluon fusion process is
written as a convolution of the partonic cross section
$\hat{\sigma}(x_g,k_{t};\; {\gamma^* g^* \to q \bar{q}})$ with
the $k_{t}$ dependent (unintegrated) gluon density ${A}(x,k_t^2,\mu^2)$.

The multidimensional integrations can be
performed by means of Monte-Carlo technique either by using
VEGAS~\cite{VEGAS} for the pure parton level calculations,
or by using the full Monte Carlo event generator \\
CASCADE~\cite{CASCADE,jung_salam_2000,CASCADEMC}.

Cross section calculations require an explicit representation of the
$k_t$ dependent (unintegrated) gluon density ${A}(x,k_t^2,\mu^2)$.   
We have used three different representations, one ({\it JB}) coming from a
leading-order perturbative solution of the BFKL
equations~\cite{Bluem}, the second set ({\it JS})
derived from a numerical solution of the CCFM
equation~\cite{CASCADE,jung_salam_2000} and the third ({\it KMR})
from solution of a
combination of the BFKL and DGLAP equations~\cite{martin_kimber}.

\section{Numerical results and discussion}

A comparison between model predictions and data in principle has to be
made on
hadron level and only if it turns out that hadronization effects are small
will a comparison to parton level predictions make sense.
However, a full simulation even of the partonic final state, including the
initial and final state QCD cascade needs a full  Monte Carlo event
generator. Such a Monte Carlo generator based on $k_t$-factorization
and using explicitly off-shell matrix elements for the hard scattering
process
convoluted with $k_t$-unintegrated gluon densities
is presently only offered by the 
CASCADE~\cite{CASCADE,jung_salam_2000,CASCADEMC}
program which uses the CCFM unintegrated gluon distribution.
\begin{figure*}[ht]
\begin{center}
\includegraphics[width=0.5\linewidth]{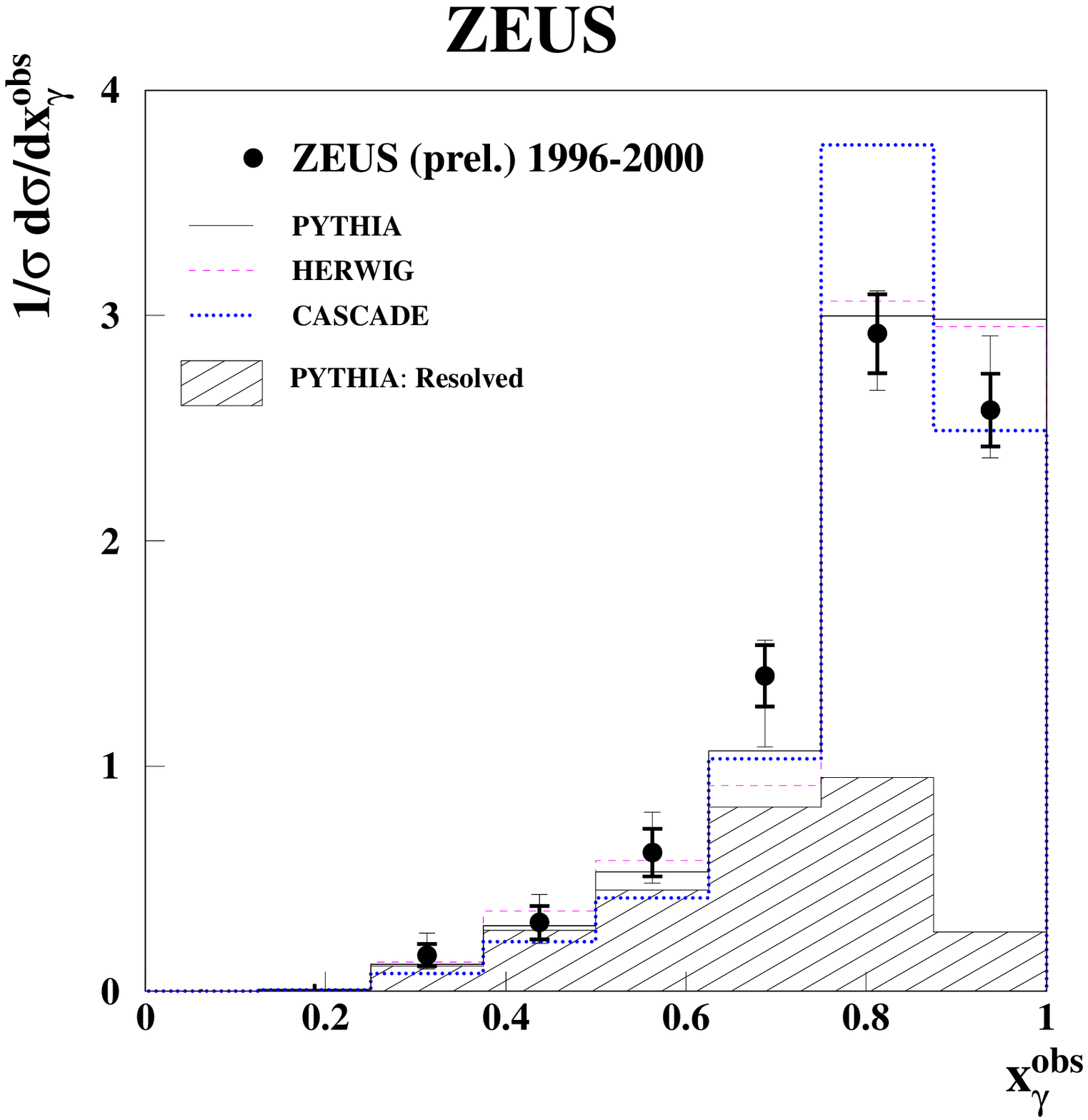}
\includegraphics[width=0.45\linewidth]{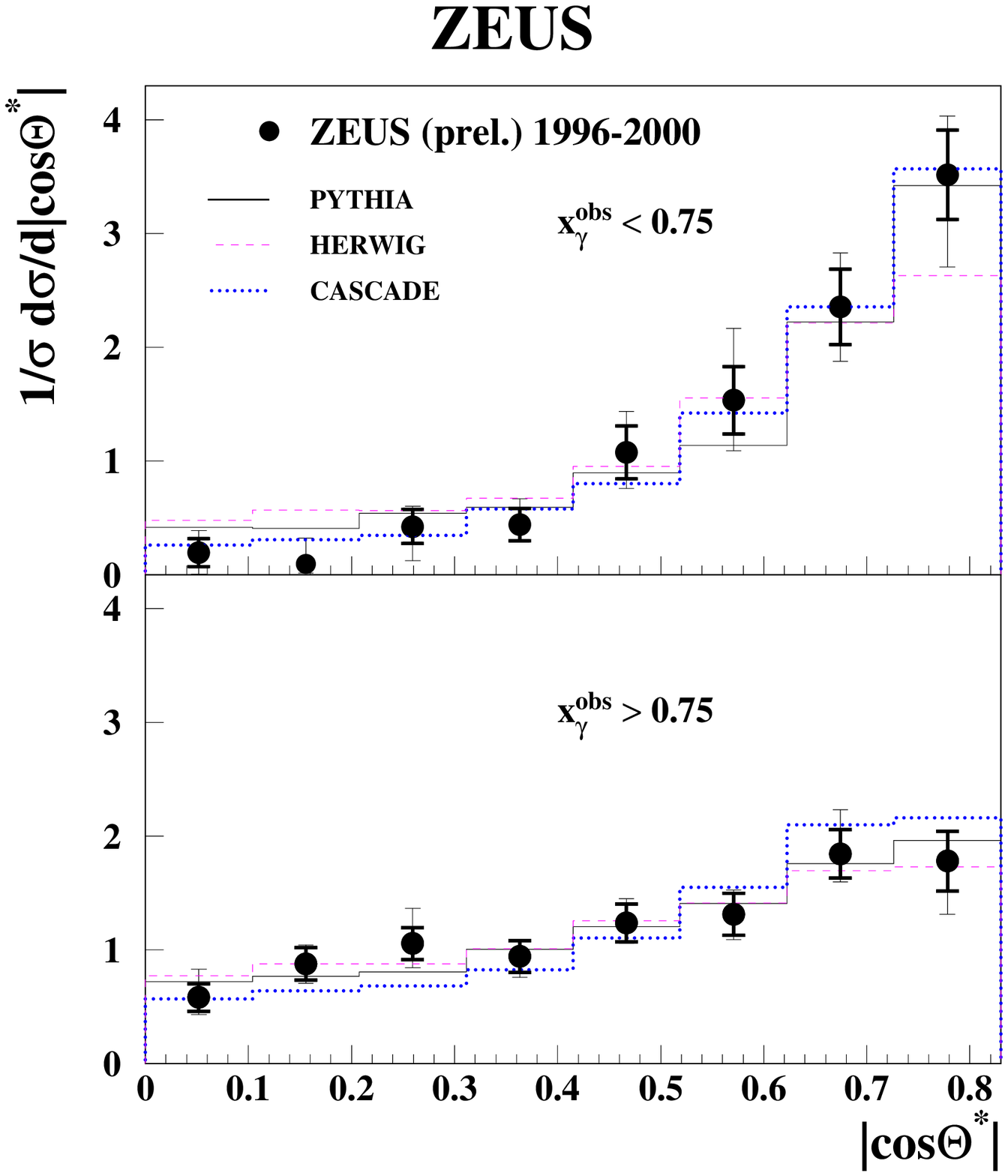}
\end{center}
\vspace*{-5mm}
\caption{{\it Relative differential cross sections  ${1/\sigma} d\sigma 
/dx_{\gamma}^{obs}$ (a) and ${1/\sigma} d\sigma/d|\cos{\theta^*}|$ (b)
with  $Q^2<1$ GeV$^2$, $p_T^{D^*} > 3$ GeV, $|{\eta}^{D^*}| < 1.5$,
$130 < W < 280$ GeV, $|{\eta}^{jet 1,2}| < 2.4$, $E_T^{jet 1,2} > 5$,
$M_{jj} > 18$ GeV and $|\bar{\eta}| < 0.7$. The histograms are 
 results with various MC simulations.}}
\label{xgamcc}
\end{figure*}

In the ref.~\cite{bjjpz} we  systematically compared the predictions
from the  $k_t$-factorization approach to
published data on charm production at HERA.
For this we use $D^*$ photo-production data from ZEUS
\cite{ZEUS-D*-gammap} and $D^*$ production in deep inelastic scattering
from both
ZEUS~\cite{ZEUS-D*-dis} and H1~\cite{H1-D*-dis}.
 First we calculate observables using a pure parton level
calculation based on the matrix element calculation of BZ~\cite{d*gam}
including the Peterson fragmentation function~\cite{Peterson}
for the transition from the charm quark to the observed $D^*$ meson.
Then we compare the result with a full hadron level simulation using the
Monte Carlo generator CASCADE  with the matrix element of CCH~\cite{CCH}.
We choose the {\it JS} unintegrated gluon for this comparison.
\par
Next we investigate on parton level different unintegrated gluon
densities. We study the sensitivity of the model predictions to the 
details of the unintegrated gluon density, the charm mass and the scale.
 We observed \cite{bjjpz}
that the $p_t$ distribution of $D^*$
mesons both in photo-production and deep inelastic scattering is in
general well described, both with the full hadron level simulation as
implemented in CASCADE and also
with the parton level calculation supplemented with the Peterson
fragmentation function. We can thus conclude, that the $p_t$ distribution
is only slightly dependent on the details of the charm fragmentation.
\par
We also consider the rapidity distribution   
of the produced $D^*$. In photo-production and in DIS the differential  
cross section
$d \sigma /d \eta$, where $\eta$ is the pseudo-rapidity of the $D^*$ 
meson, is
sensitive to the choice of the unintegrated gluon distribution,
 We observed, that the parton level prediction
including the Peterson fragmentation function is not able to describe
the measurement over the full range of $\eta$. The effect of a full
hadron level simulation is clearly visible as CASCADE provides a much
better description of the experimental data.

Then we investigate the $x_{\gamma}$ distribution,
which is sensitive to the details of the initial state cascade.
We compare the predictions from a pure parton level calculation and a full
event simulation of CASCADE with the measurements (Fig. 1(a)).
 We can
conclude that the $k_t$-factorization approach effectively simulates heavy
quark excitation and indeed the hardest $p_t$ emission comes frequently
from a gluon in the initial state cascade~\cite{xgam}.

Other interesting quantities are the dijet angular distributions of 
resolved  
photon like events ($x^{OBS}_{\gamma} < 0.75$)
compared to the direct-photon  like
events ($x^{OBS}_{\gamma} > 0.75$).
In the $k_t$-factorization approach the angular distribution will be
determined from the off-shell matrix element, which covers both scattering 
processes.
 Comparisions of the CASCADE results with the ZEUS experimental data
for these angular distributions were done  by S. Padhi
 \cite{padhi} (Fig. 1(b)).


In summary we have shown, that the $k_t$ - factorization approach can be 
consistently
used
to describe measurements of charm production at HERA, which are
known to be not well reproduced
in the collinear approach. We have also shown, that in  
$k_t$-factorization, resolved photon like processes are effectively  
simulated
including the proper angular distributions.

\end{document}